\documentclass[aps,prb,floatfix,showpacs,twocolumn]{revtex4}
\usepackage{amsmath,amssymb}
\usepackage{epsfig}

\newcommand{\bm}{\boldsymbol}

\begin{document}

%\twocolumn[
\hsize\textwidth\columnwidth\hsize\csname@twocolumnfalse\endcsname

\title{Coulomb Drag and Spin Coulomb Drag in the presence of Spin-orbit Coupling}

\author{Wang-Kong Tse}
\author{S. Das Sarma}
\affiliation{Condensed Matter Theory Center, Department of Physics,
University of Maryland, College Park, Maryland 20742}

\begin{abstract}
Employing diagrammatic perturbation theory, we calculate
the (charge) Coulomb drag resistivity $\rho_D$ and spin Coulomb drag
resistivity $\rho_{\uparrow\downarrow}$ in the presence of Rashba
spin-orbit coupling. Analytical expressions for $\rho_D$ and
$\rho_{\uparrow\downarrow}$ are derived, and it is found that
spin-orbit interaction produces a small enhancement to $\rho_D$ and
$\rho_{\uparrow\downarrow}$ in the ballistic regime while $\rho_D$
is unchanged in the diffusive regime. This enhancement in the ballistic regime is attributed to the enhancement of
the nonlinear susceptibility (i.e. current produced through the 
rectification of the thermal electric potential fluctuations in the
passive layer) while the lack of enhancement in the diffusive regime
is due to the suppression by disorder.

\end{abstract}
%]
%{73.43.-f, 72.25.Dc, 75.80.+q, 71.70.Ej}

\maketitle
\newpage

\section{Introduction}

Recent years have seen the emergence of a strong interest in exploiting the
manipulation of the spin degrees of freedom in solid state systems
that could lead to enhanced performance in electronic devices;
this field of spin electronics,
or simply spintronics \cite{RMP}, has evolved into an exciting subject in condensed matter physics.
Among the myriad of spintronics proposals, gate-controlled Rashba
spin-orbit (SO) coupling has attracted tremendous attention as it
offers the possibility of all-electrical spin manipulation without
the presence of a magnetic field. One such experimentally realizable
 heterostructure is a double quantum well gate-modulated with Rashba spin-orbit
coupling. In this paper, we study theoretically the effect of 
spin-orbit coupling on the Coulomb drag properties of 2D bilayer
systems. As is well known, there exists Coulomb coupling between the
barrier-separated two-dimensional electron gas (2DEG) layers induced
by the momentum exchanges of the Coulomb-scattered electrons residing
in individual layers. This phenomenon of Coulomb drag (in the absence
of spin-orbit coupling), first observed experimentally by Gramila \textit{et
  al}. \cite{Gramila} between two 2DEG layers, has continued to be a
subject of thorough investigation \cite{DragReview}; with a theoretical description
developed in Refs.~\cite{MacDonald,Jauho2,Oreg,Jauho}. 
The physical mechanism at work in Coulomb drag can be understood as follows: The
application of a current through one layer (the active layer) causes
thermal fluctuations in the electron density, which are transferred
across the barrier to the other layer (the passive layer) due to
momentum exchanges by electron-electron scattering. In the passive
layer, the variations in electrical potential associated with
the thermal fluctuations in electron density then produce a current
due to second-order rectification effect, quantified by the
nonlinear susceptibility which is the response function connecting
the random potential fluctuations and the induced electrical current.
The drag resistivity $\rho_D$, which is defined as $\rho_D =
E_{\mathrm{passive}}/j_{\mathrm{active}}$ --- the ratio of the
electric field strength developed in the open-circuited passive layer to the current density
in the active layer, is a useful experimental measure of the Coulomb
drag phenomenon.

Although the subject of Coulomb drag has existed since the first
experiment more than a decade ago, it remains a 
very vibrant research topic today: recent experimental \cite{HoleD1}
and theoretical \cite{HoleD2} studies have revealed that bilayer hole
drag in the low density regime is a factor of $\sim 10^2-10^3$
larger than the corresponding electron drag. Moreover, other more recent
experiments \cite{HoleD3,HoleD5} have shown that the low-density
bilayer hole drag in the presence of an in-plane magnetic field
(i.e. magnetodrag) has a qualitative dependence
on the applied magnetic field unexpectedly similar to that of the 
single-layer magnetoresistivity, and this surprising observation has been 
accounted for theoretically by properly considering the suppression of
screening by carrier spin polarization within the framework of the
standard Fermi-liquid theory \cite{HoleD4}. In light of these recent
developments, a very interesting question is whether
spin-orbit coupling (which is effectively a momentum-dependent in-plane
magnetic field) would change the qualitative behaviour of the drag
resistivity. To our knowledge, the effect of spin-orbit coupling on Coulomb drag
has not yet been investigated, and this problem is highly relevant in
view of the current growing
interest in viable semiconductor spintronic systems and especially potential
device applications employing gate-controlled spin-orbit coupling.

The bilayer Coulomb drag provides direct quantitative information
about carrier interaction effects since the drag voltage in the
passive layer arises entirely due to the carrier-carrier scattering
between the layers. The study of bilayer Coulomb drag properties in
the presence of spin-orbit coupling effects therefore introduces a
method of investigating the interplay between spin-orbit coupling and
interaction effects. Since both spin-orbit coupling and interaction
can be tuned by applying external electric and magnetic fields
respectively, and also by changing the 2D density, the possibility
exists for a detailed understanding of the interplay between
spin-orbit coupling and many-body interaction effects through bilayer
drag studies. This is a main motivation for our theoretical work.

On the other hand, it has also been proposed that there exists an
analogous effect, called the spin Coulomb drag \cite{Vignale1}, in
which spin-up and spin-down electrons within one single device structure (three- or
two-dimensional) play the roles of the electrons in each individual
layers in the charge Coulomb drag problem, and now it is the Coulomb
scattering between the spin-up and spin-down carriers that causes the
frictional force, damping the relative motion of the two spin
components. Therefore, unlike ordinary charge current, the flow of
opposite spin carriers (i.e. spin current) is a non-conserved
quantity even without spin-orbit coupling and tends to be suppressed
because of this drag effect. In the same spirit as $\rho_D$, one gains
a measure of the spin drag by defining $\rho_{\uparrow\downarrow} =
E_{\uparrow}/j_{\downarrow}$, the spin drag resistivity given by the ratio of the gradient
of the electrochemical potential of spin-up carriers to the current
density of the spin-down carriers, with the current of the spin-up
carriers held zero. This effect has been confirmed in a recent experimental paper by
Weber \textit{et al}. \cite{Weber} where it was found that the
suppression of the measured spin diffusion coefficient relative to the charge diffusion
coefficient could be explained by a correction factor
$1/(1+\vert\rho_{\uparrow\downarrow}\vert/\rho)$ coming from the
theoretical spin drag prediction. We note that while the Coulomb drag
refers to a bilayer system, spin drag refers to a single 2D layer (or
a single 3D system) with the up and down components of the spin playing the role of the
``active'' and ``passive'' carrier components.

The purpose of the present paper is to investigate the effect of spin-orbit
interaction on the drag phenomenon in semiconductor structures
including the usual (charge) Coulomb drag and the spin Coulomb
drag. We follow Refs.~\cite{Oreg,Jauho} and apply diagrammatic perturbation theory to
calculate the Coulomb drag resistivity $\rho_D$ and spin Coulomb drag
resistivity $\rho_{\uparrow\downarrow}$ in the presence of Rashba SO coupling. We find that for
clean samples the Coulomb drag resistivity and the spin drag
resistivity are enhanced by a correction factor
$\Delta\rho_{D,\uparrow\downarrow}/\rho_{D,\uparrow\downarrow} \sim
\mathcal{O}(\gamma^2)$ in the presence of SO coupling ($\gamma$ is the
dimensionless spin-orbit coupling strength), whereas for
dirty samples the spin-orbit coupling correction to the Coulomb drag
is essentially completely suppressed by disorder.

Our paper is organized as follows: we begin in section II with the
problem of Coulomb drag,
where after recapitulating a set of useful formulas related to Rashba SO
coupling and Coulomb drag we proceed to evaluate in both the
ballistic limit (subsection A) and the diffusive limit (subsection
B) the nonlinear susceptibility, the central quantity in the
problem. These results are then used in subsection C to
calculate the Coulomb drag resistivity in the presence of Rashba SO
coupling. With the aid of the formalism applied to the Coulomb drag,
we turn in section III to the problem of spin Coulomb drag and calculate the spin drag
resistivity in the ballistic regime. Finally conclusions are presented in section IV.

\section{Coulomb Drag}

The electronic Hamiltonian in the presence of the Rashba SO coupling
\begin{equation}
H = \frac{p^2}{2m}+\alpha\left(\bm{\sigma}\times\bm{p}\right)\cdot\hat{\bm{z}},
\label{eq1}
\end{equation}
has the following eigenstates
\begin{eqnarray}
\vert\bm{p}\mu\rangle = \frac{1}{\sqrt{2}}
\left[
\begin{array}{c}
1 \\
\mu ie^{i\phi}
\end{array}
\right]\vert\bm{p}\rangle, \label{eq2}
\end{eqnarray}
which is labelled by the quantum number $\mu = \pm 1$ called the chirality. The corresponding
eigenenergy spectra follow as $\epsilon_{\bm{p}\pm} = p^2/2m \mp \alpha
p$. We note that the transformation which diagonalizes the Hamiltonian
Eq.~(\ref{eq1}) is a local transformation, dependent on the momentum
$\bm{p}$ through $\mathrm{tan}\phi = p_y/p_x$:
\begin{eqnarray}
U_{\bm{p}} = \frac{1}{\sqrt{2}}
\left[
\begin{array}{cc}
1 & 1 \\
ie^{i\phi} & -ie^{i\phi}
\end{array}
\right]. \label{eq3}
\end{eqnarray}
In the presence of Rashba SO coupling, the Fermi energy $\varepsilon_F
= \hbar^2 k_F^2/2m$ of the system is lowered in comparison to the case
without SO coupling for the same number of electrons and there exists two branches of Fermi wavevectors
$k_{F\pm}$ corresponding to the two chiralities. From
$\epsilon_{\bm{k}}\mp\alpha \hbar k -\varepsilon_F = 0$ we can express the
two solutions $k_{\pm}$ in terms of $k_F$; on the other hand by particle
conservation we have $n_{+}+n_{-} = k_{+}^2/4\pi+k_{-}^2/4\pi =
k_F^2/2\pi$. Solving then yields the Fermi energy and Fermi wavevectors in
the presence of Rashba SO coupling:
\begin{eqnarray}
\varepsilon_F &=& \varepsilon_{F0}\left(1-2\gamma^2\right), \\ \label{eq41}
k_{F\pm} &=& k_{F0}\left(\pm \gamma+\sqrt{1-\gamma^2}\right), \label{eq42}
\end{eqnarray}
where $\varepsilon_{F0}$ and $k_{F0}$ are the Fermi energy and Fermi
wavevector without SO coupling, here we also define the dimensionless 
spin-orbit coupling strength $\gamma = \alpha/v_{F0} \ll 1$. The density of
states $\nu$ at the Fermi level is readily found as
\begin{equation}
\nu_{\pm} = \nu\left(1\pm\frac{\gamma}{\sqrt{1-\gamma^2}}\right), \label{eq43}
\end{equation}
where $\nu = m/2\pi \hbar^2$ is the density
of states per spin. In the following we proceed to evaluate the Coulomb drag resistivity
of a double-layer 2DEG system with
Rashba spin-orbit coupling. Using linear response theory, the
drag resistivity is obtained from the diagrammatic expansions in
Fig.~1 and is given by \cite{Oreg,Jauho}:
\begin{eqnarray}
\rho_{\mathrm{D}} &=& \frac{\hbar}{16\pi k_B T\sigma_1\sigma_2}\sum_{\bm{q}}\int_{0}^{\infty}
\frac{\mathrm{d}\omega}{\mathrm{sinh}^2\left(\hbar\omega/2k_B
    T\right)} \nonumber \\
&&\Gamma_{1x}\left(q,\omega\right)\Gamma_{2x}\left(q,\omega\right)\left\vert U_{12}\left(q,\omega\right)\right\vert^2.
\label{eq4}
\end{eqnarray}
\begin{figure}
  \includegraphics[width=4.5cm,angle=0]{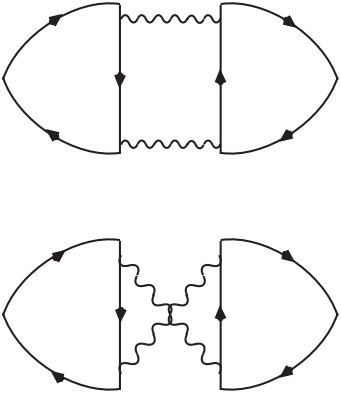}
\caption{Diagrams contributing to the drag resistivity
  Eq.~(\ref{eq4}). For Coulomb drag (spin Coulomb drag), the wavy lines
  represent interlayer (intralayer) Coulomb potential; the vertices on
  the left and on the right denote charge current in layer 1 and layer
  2 (spin-up current and spin-down current) respectively.} \label{fig1}
\end{figure}
\begin{figure}
  \includegraphics[width=6.5cm,angle=0]{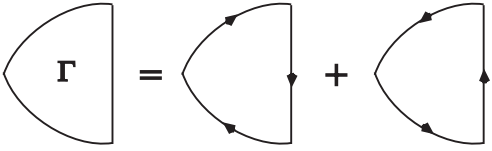}
\caption{The three-point vertex representing the nonlinear
  susceptibility $\Gamma$. The vertex on the right side represent
  current vertex while the two vertices at the top and the bottom
  represent charge density vertices.} \label{fig2}
\end{figure}
where $\sigma = 2e^2 \nu D$ is the Boltzmann conductivity of the
individual layer ($D = \varepsilon_{F0}\tau/m$ is the diffusion constant), the subscripts `1',`2' signify quantities in
layer 1 and 2 respectively. $U_{12}$ is the screened interlayer potential
which is obtained from solving the corresponding Dyson equation in the
random phase approximation (RPA), $\bm{\Gamma}$ is the
nonlinear susceptibility and is given from the
three-point vertex diagrams in Fig.~2 as 

\begin{eqnarray}
&&\bm{\Gamma}(\bm{q},\omega) = \frac{1}{2\pi i}\int \mathrm{d}\varepsilon
\left[n_F(\varepsilon+\omega)-n_F(\varepsilon)\right]\sum_p
\mathrm{tr} \nonumber \\
&&\;\left\{\left[G_{\varepsilon}^{A}(\bm{p}-\bm{q})-G_{\varepsilon}^{R}(\bm{p}-\bm{q})\right]
G_{\varepsilon+\omega}^{A}(\bm{p})\bm{J}(\bm{p})G_{\varepsilon+\omega}^{R}(\bm{p})\right\}
\nonumber \\
&&+\left\{\bm{q},\omega \to -\bm{q},-\omega\right\},
\label{eq5}
\end{eqnarray}
where $G_{\varepsilon}^{R,A}(\bm{p}) = (\varepsilon-H \pm i\hbar/2\tau)^{-1}$
denotes the retarded/advanced Green function, and $\mathrm{tr}$ the trace. Note that in the expression of the drag resistivity Eq.~(\ref{eq4}),
because of the denomenator $\mathrm{sinh}^2(\hbar \omega/2k_B T)$, the
dominant contribution of the $\varepsilon$ integral comes from the region $\hbar \omega \lesssim k_B
T$, and therefore at low temperatures the dominant contribution to the
drag resistivity comes from small values of $\omega$. In the following sections we evaluate Eq.~(\ref{eq5}) in respect to
two regimes: ballistic regime ($\omega\tau > 1$ or $ql > 1$) and
diffusive regime ($\omega\tau, ql \ll 1$), where $l = v_{F0}\tau$ is the mean
free path. \\

\subsection{Ballistic limit}

In the ensuing discussions we express all Green functions and currents in the chiral
basis, making use of the transformation Eq.~(\ref{eq3}). To this end
we define the transformation $U_{+} =
U_{\bm{p}}^{\dag}U_{\bm{p}+\bm{q}}$ and $U_{-} = U_{+}^{\dag} =
U_{\bm{p}}^{\dag}U_{\bm{p}-\bm{q}}$; explicitly,
\begin{eqnarray}
U_{+} = \frac{1}{{2}}
\left[
\begin{array}{cc}
1+\mathrm{e}^{i\theta} & 1-\mathrm{e}^{i\theta} \\
1-\mathrm{e}^{i\theta} & 1+\mathrm{e}^{i\theta}
\end{array}
\right], \label{eq7}
\end{eqnarray}
where $\theta = \phi_{\bm{p}+\bm{q}}-\phi_{\bm{p}} =
\phi_{\bm{p}}-\phi_{\bm{p}-\bm{q}}$ is the scattering angle from
momentum $\bm{p}$ to $\bm{p}+\bm{q}$ (or equivalently from momentum
$\bm{p}-\bm{q}$ to $\bm{p}$). It will be useful to also note that $U_{+,-}^{\dag} =
U_{+,-}^{*}$ since $U_{+,-}$ is a symmetric matrix.

After a change of variable in $\varepsilon$ in the
second term and a transformation into the chiral basis,
Eq.~(\ref{eq5}) can be written as
\begin{widetext}
\begin{eqnarray}
\bm{\Gamma}(\bm{q},\omega) &=& \frac{1}{2\pi i}\int \mathrm{d}\varepsilon
\left[n_F(\varepsilon+\omega)-n_F(\varepsilon)\right]
\sum_p
\mathrm{tr} \left\{U_{-}
\left[\tilde{G}_{\varepsilon}^{A}(\bm{p}-\bm{q})-\tilde{G}_{\varepsilon}^{R}(\bm{p}-\bm{q})\right]U_{-}^{\dag}
\tilde{G}_{\varepsilon+\omega}^{A}(\bm{p})\tilde{\bm{J}}(\bm{p})\tilde{G}_{\varepsilon+\omega}^{R}(\bm{p})
\right.\nonumber \\
&&\left.-U_{+}
\left[\tilde{G}_{\varepsilon+\omega}^{A}(\bm{p}+\bm{q})-\tilde{G}_{\varepsilon+\omega}^{R}(\bm{p}+\bm{q})\right]U_{+}^{\dag}
\tilde{G}_{\varepsilon}^{A}(\bm{p})\tilde{\bm{J}}(\bm{p})\tilde{G}_{\varepsilon}^{R}(\bm{p})\right\}, \label{eq19}
\end{eqnarray}
\end{widetext}
where the tilda represents quantities expressed in the chiral basis. In the ballistic limit, the charge current in the chiral basis is given by
\begin{eqnarray}
\tilde{J}_x &=& \frac{\hbar k_x}{m}-\alpha \tilde{\sigma}_y \nonumber \\
&=& \left[
\begin{array}{cc}
\hbar k_x/m-\alpha \mathrm{cos}\phi & i\alpha\mathrm{sin}\phi \\
- i\alpha\mathrm{sin}\phi & \hbar k_x/m+\alpha \mathrm{cos}\phi
\end{array}
\right]. \label{eq20}
\end{eqnarray}
\\
In the following we will denote the matrix elements of $\tilde{J}_x$
as $\tilde{J}_{ij}$ for brevity. Here it will be useful to note that
the off-diagonal elements are related by the hermiticity of
$\tilde{\bm{J}}$ as $\tilde{J}_{12} = \tilde{J}_{21}^{*} = -\tilde{J}_{21}$.

Consider the energy integration of the first term in Eq.~(\ref{eq19}):
\begin{widetext}
\begin{eqnarray}
&&\int \mathrm{d}\varepsilon\left[n_F(\varepsilon+\omega)-n_F(\varepsilon)\right] \mathrm{tr}\left\{U_{-}
\left[\tilde{G}_{\varepsilon}^{A}(\bm{p}-\bm{q})-\tilde{G}_{\varepsilon}^{R}(\bm{p}-\bm{q})\right]U_{-}^{\dag}
\tilde{G}_{\varepsilon+\omega}^{A}(\bm{p})\tilde{\bm{J}}(\bm{p})\tilde{G}_{\varepsilon+\omega}^{R}(\bm{p})\right\}
\nonumber \\
&=&
i\frac{2\pi\tau}{\hbar}\left\{\tilde{J}_{11}\left[\left(1+\mathrm{cos}\theta\right)\mathrm{Im}\left\{\frac{n_F(\epsilon_{\bm{p-q}+})-n_F(\epsilon_{\bm{p}+})}{\omega+\epsilon_{\bm{p-q}+}-\epsilon_{\bm{p}+}+i0^{+}}\right\}
+\left(1-\mathrm{cos}\theta\right)\mathrm{Im}\left\{\frac{n_F(\epsilon_{\bm{p-q}-})-n_F(\epsilon_{\bm{p}+})}{\omega+\epsilon_{\bm{p-q}-}-\epsilon_{\bm{p}+}+i0^{+}}\right\}\right] \right.
\nonumber \\
&&\;\;\;\;\;\;\;\;\,+\tilde{J}_{22}\left[\left(1+\mathrm{cos}\theta\right)\mathrm{Im}\left\{\frac{n_F(\epsilon_{\bm{p-q}-})-n_F(\epsilon_{\bm{p}-})}{\omega+\epsilon_{\bm{p-q}-}-\epsilon_{\bm{p}-}+i0^{+}}\right\}
+\left(1+\mathrm{cos}\theta\right)\mathrm{Im}\left\{\frac{n_F(\epsilon_{\bm{p-q}+})-n_F(\epsilon_{\bm{p}-})}{\omega+\epsilon_{\bm{p-q}+}-\epsilon_{\bm{p}-}+i0^{+}}\right\}\right]
\nonumber \\
&&\left.\;\;\;\;\;\;\;\;+2\pi \tilde{J}_{12}\mathrm{sin}\theta
\mathrm{Im}\left\{\frac{n_F(\epsilon_{\bm{p-q}+})-n_F(\epsilon_{\bm{p}+})}{\omega+\epsilon_{\bm{p-q}+}-\epsilon_{\bm{p}+}+i0^{+}}-\frac{n_F(\epsilon_{\bm{p-q}-})-n_F(\epsilon_{\bm{p}-})}{\omega+\epsilon_{\bm{p-q}-}-\epsilon_{\bm{p}-}+i0^{+}}\right\}\delta\left(\epsilon_{\bm{p}+}-\epsilon_{\bm{p}-}\right)\right\}.
\label{eq21}
\end{eqnarray}
\end{widetext}
We notice that the last term in Eq.~(\ref{eq21}) contains a delta
function $\delta\left(\epsilon_{\bm{p}+}-\epsilon_{\bm{p}-}\right) =
\delta\left(2\alpha p\right)$, which forces this term to vanish
identically upon performing the momentum integration in
Eq.~(\ref{eq19}), and hence, rather unexpectedly, the off-diagonal 
terms in the current Eq.~(\ref{eq20}) drop out and
only the diagonal terms contribute to the nonlinear susceptibility.
The same analysis above follows for the second term in
Eq.~(\ref{eq19}), and the expression for the nonlinear
susceptibility reduces to (with the notation $\tilde{J}_{++,--} =
\tilde{J}_{11,22}$):
\begin{eqnarray}
&&{\Gamma}_x(\bm{q},\omega) = \frac{\tau}{\hbar}
\sum_{\mu,\mu'=\pm}\sum_{\bm{p}}
\tilde{J}_{\mu\mu}(\bm{p})\left(1+\mu\mu'\mathrm{cos}\theta\right) \label{eq22} \\
&&\mathrm{Im}\left\{\frac{n_F(\epsilon_{\bm{p-q}\mu'})-n_F(\epsilon_{\bm{p}\mu})}{\omega+\epsilon_{\bm{p-q}\mu'}-\epsilon_{\bm{p}\mu}+i0^{+}}
-\frac{n_F(\epsilon_{\bm{p}\mu'})-n_F(\epsilon_{\bm{p+q}\mu})}{\omega+\epsilon_{\bm{p}\mu'}-\epsilon_{\bm{p+q}\mu}+i0^{+}}
\right\}. \nonumber \\
%\nonumber \\
%&&=\frac{\tau}{\hbar}\sum_{\mu,\mu' = \pm} \sum_{\bm{p}}
%\left[\tilde{J}_{\mu\mu}(\bm{p+q})-\tilde{J}_{\mu\mu}(\bm{p})\right]\left(1+\mu\mu'\mathrm{cos}\theta\right)
%\nonumber \\
%&&\;\;\;\;\;\;\;\;\;\;\;\;\;\;\;\;\;\;\;\;\;\;\;\;\;\mathrm{Im}\left\{\frac{n_F(\epsilon_{\bm{p}\mu'})-n_F(\epsilon_{\bm{p+q}\mu})}{\omega+\epsilon_{\bm{p}\mu'}-\epsilon_{\bm{p+q}\mu}+i0^{+}}\right\}.
\nonumber
\end{eqnarray}

Up to this point we have not yet made use of any simplifying assumptions and Eq.~(\ref{eq22})
for the ballistic case is general. Now, we evaluate Eq.~(\ref{eq22})
in the limit of low-energy  $u \ll 1$ and long-wavelength $x \ll 1$,
and we neglect the interband
terms ($\mu \ne \mu'$) as these are readily shown to be smaller than their intraband
counterparts ($\mu = \mu'$) by an order of
$\mathcal{O}(q^2/k_{F0}^2)$. At low temperatures $k_B T \ll
\varepsilon_{F0}$, we expand the Fermi functions in $\varepsilon$ in Eq.~(\ref{eq22}) and keep up to
$\mathcal{O}(\omega)$. The evaluation of a typical intraband term then goes as
\begin{eqnarray}
&&\sum_{\bm{p}}
\tilde{J}_{11}(\bm{p})\left(1+\mathrm{cos}\theta\right)\mathrm{Im}\left\{\frac{n_F(\epsilon_{\bm{p-q}+})-n_F(\epsilon_{\bm{p}+})}{\omega+\epsilon_{\bm{p-q}+}-\epsilon_{\bm{p}+}+i0^{+}}\right\}
\nonumber \\
%&\simeq&
%-2\pi\omega\sum_{\bm{p}}\tilde{J}_{11}(\bm{p})\delta\left(\epsilon_{\bm{p}+}-\epsilon_F\right)
%\delta\left(\omega+\epsilon_{\bm{p}+}-\epsilon_{\bm{p+q}+}\right)
%\nonumber \\
%&=& -2\omega\frac{2m}{\hbar^2
%  k_{F0}}\frac{1}{\sqrt{1-\gamma^2}}\sum_{\bm{p}}\tilde{J}_{11}(\bm{p})\delta\left(k-k_{F+}\right) \nonumber \\
%&&\delta\left[\omega+\epsilon_{\bm{q}}-\frac{\hbar}{m}\left(k-\gamma k_{F0}\right)q\mathrm{cos}\phi\right] \nonumber \\
&=&
-\frac{2uk_{F0}}{\pi^2\hbar}\left(1+\frac{\gamma}{\sqrt{1-\gamma^2}}\right)\frac{u+x}{\sqrt{1-\gamma^2}},
\label{eq24}
\end{eqnarray}
where $u = \omega/v_{F0} q$ and $x = q/2k_{F0}$, $v_{F0} = \hbar k_{F0}/m$ is the Fermi velocity without
spin-orbit coupling. Evaluating the rest of the intraband terms in the same manner, we find the nonlinear
susceptibility Eq.~(\ref{eq22}) as
\begin{eqnarray}
{\Gamma}_x(q,\omega) &=&-\frac{4\nu Dq}{\varepsilon_{F0}}\frac{u}{\sqrt{1-\gamma^2}}
\nonumber \\
&\simeq&-\frac{4\nu Dq}{\varepsilon_{F0}}{u}\left(1+\frac{1}{2}\gamma^2\right).
\label{eq25}
\end{eqnarray}
Here we notice that the nonlinear susceptibility $\Gamma(q,\omega)$ is
enhanced in the presence of SO coupling. In the ballistic limit, it
should be emphasized that the nonlinear susceptibility
without SO coupling is directly proportional to the
imaginary part of the irreducible polarizability generally \cite{remark,Jauho}, independent of the low-energy long-wavelength assumption
made on $u$ and $x$. Such a general statement is not true 
with SO coupling present, where it is readily seen that Eq.~(\ref{eq22}) is not, despite the similarity,
explicitly proportional to the imaginary
part of the irreducible polarizability in the presence of Rashba SO
coupling \cite{saraga}:
\begin{eqnarray}
&&\Pi\left(q,\omega\right) = \label{eq23} \\
&&-\frac{1}{2}\sum_{\mu,\mu' = \pm}\sum_{\bm{p}}
\left(1+\mu\mu'\mathrm{cos}\theta\right)\frac{n_F(\epsilon_{\bm{p}\mu'})-n_F(\epsilon_{\bm{p+q}\mu})}{\omega+\epsilon_{\bm{p}\mu'}-\epsilon_{\bm{p+q}\mu}+i0^{+}}.
\nonumber
\end{eqnarray}
The reason here is that the current term
$\tilde{J}_{11,22}(\bm{p+q})-\tilde{J}_{11,22}(\bm{p})$, with SO
coupling (c.f. Eq.~(\ref{eq20})), is no longer simply
proportional to $q$ and so cannot be taken outside of the momentum
integral. In the long-wavelength low-energy limit, however,
proportionality is restored as can be seen by comparing with the following expression
of the irreducible polarizability evaluated in the same limit:
\begin{equation}
\Pi\left(q,\omega\right) \simeq 2\nu \left(1+i\frac{u}{\sqrt{1-\gamma^2}}\right).
\label{eq27}
\end{equation}
Therefore, in the ballistic regime, the enhancement of the nonlinear susceptibility Eq.~(\ref{eq25}) is
physically attributable to the SO-induced enhancement of the intraband
electron-electron scattering.

\subsection{Diffusive limit}

For low temperatures $k_B T \ll \varepsilon_{F0}$, we expand the
Fermi functions in $\varepsilon$ in Eq.~(\ref{eq5}) and keep up to
$\mathcal{O}(\omega)$. Transforming into the chiral basis, $\Gamma$
can then be expressed as
\begin{eqnarray}
&&\bm{\Gamma}(\bm{q},\omega) = -\frac{\omega}{2\pi i}\sum_p
\mathrm{tr} \label{eq8} \\
&&\left\{U_{-}
\left[\tilde{G}_{\varepsilon_F}^{A}(\bm{p}-\bm{q})-\tilde{G}_{\varepsilon_F}^{R}(\bm{p}-\bm{q})\right]U_{-}^{\dag}
\right.\nonumber \\
&&\left.\;\;\tilde{G}_{\varepsilon_F+\omega}^{A}(\bm{p})\tilde{\bm{J}}(\bm{p})\tilde{G}_{\varepsilon_F+\omega}^{R}(\bm{p})\right\}
+\left\{\bm{q},\omega \to -\bm{q},-\omega\right\}.
%&&-\left\{U_{+}
%\left[\tilde{G}_{\varepsilon_F}^{A}(\bm{p}+\bm{q})-\tilde{G}_{\varepsilon_F}^{R}(\bm{p}+\bm{q})\right]U_{+}^{\dag}
%\tilde{G}_{\varepsilon_F-\omega}^{A}(\bm{p})\tilde{\bm{J}}(\bm{p})\tilde{G}_{\varepsilon_F-\omega}^{R}(\bm{p})\right\},
\nonumber
\end{eqnarray}
Since the spin-orbit coupling strength is assumed to be weak $\gamma \ll 1$, we retain terms
up to the first nonvanishing order of $\mathcal{O}(\gamma^2)$. In the diffusive limit
$\omega \ll 1/\tau, q \ll 1/l$, we also expand Eq.~(\ref{eq8}) in terms of these
parameters up to the first nonvanishing order, e.g. for the Green function
we have
\begin{eqnarray}
\tilde{G}_{\varepsilon_F}(\bm{p}\pm\bm{q}) \simeq
\tilde{G}_{\varepsilon_F}(\bm{p})\pm \bm{q}\cdot\hat{\bm{p}}\left(\frac{p}{m}-\alpha\sigma_z\right)\left[\tilde{G}_{\varepsilon_F}(\bm{p})\right]^2.
\label{eq9}
\end{eqnarray}
For the charge current vertex in the Rashba model, taking account of
the diffuson pole
correction exactly cancels the spin-dependent term in the charge current, leaving only the bare
current term $J_x(\bm{p}) = \hbar k_x/m$.
%$\bm{p}/m$,
%
%\begin{equation}
%J_x(\bm{p}) = \hbar k_x/m.
%\label{eq10}
%\end{equation}
%
Using $U_{+,-}^{\dag} = U_{+,-}^{*}$, the nonlinear susceptibility can be
written as (dropping the energy and momentum labels of the Green
function which are understood here as $\bm{p}$ and $\varepsilon_F$):
\begin{widetext}
\begin{eqnarray}
\Gamma_x(\bm{q},\omega) = -\frac{\omega}{\pi i}\sum_p \frac{p_x}{m}
\mathrm{tr}\left\{\left\{\mathrm{Re}\left\{U_{+}
\left[\tilde{G}^{R}-\tilde{G}^{A}\right]U_{+}^{*}\right\}%\right.\right.\nonumber
+i \bm{q}\cdot\hat{\bm{p}}
\mathrm{Im}\left\{U_{+}\left(\frac{p}{m}-\alpha\sigma_z\right)\left[(\tilde{G}^{R})^2-(\tilde{G}^{A})^2\right]U_{+}^{*}\right\}\right\}\tilde{G}^{A}\tilde{G}^{R}\right\}.
\nonumber \\
\label{eq11}
%&&\left.\left.-i \bm{q}\cdot\hat{\bm{p}} \alpha\mathrm{Im}\left\{U_{+}\sigma_z\left[(\tilde{G}^{R})^2-(\tilde{G}^{A})^2\right]U_{+}^{*}\right\}
%\tilde{G}^{A}\tilde{G}^{R}\right\}. \label{eq11}
\end{eqnarray}
\end{widetext}
The first term inside the summation is readily shown to be
identically zero after the trace is taken, in which case we simply have
\begin{eqnarray}
&&\Gamma_x({q},\omega) = -\frac{\omega q}{\pi}\sum_p \mathrm{cos}^2 \phi_p\left(\frac{p}{m}\right)
\mathrm{tr} \label{eq12} \\
&&\left\{\mathrm{Im}\left\{U_{+}\left(\frac{p}{m}-\alpha\sigma_z\right)\left[(\tilde{G}^{R})^2-(\tilde{G}^{A})^2\right]U_{+}^{*}\right\}\tilde{G}^{A}\tilde{G}^{R}\right\}.
\nonumber
%&&\left.\left.-\alpha\mathrm{Im}\left\{U_{+}\sigma_z\left[(\tilde{G}^{R})^2-(\tilde{G}^{A})^2\right]U_{+}^{*}\right\}\right\}\tilde{G}^{A}\tilde{G}^{R}\right\}.
\end{eqnarray}
We proceed to evaluate the integral in Eq.~(\ref{eq12}) in the following. In the
 transformation $U_{+}\cdots U_{+}^{\dag}$, we expand and retain terms up to the
 first order in $q/k_{F0}$, i.e. $1\pm\mathrm{cos}\theta \simeq
 1\pm(1-q^2\mathrm{sin}\phi/2k^2)$ and $\mathrm{sin}\theta \simeq q\mathrm{sin}\phi/k$.
Physically, this means taking
 account only of the intraband contribution 
%$1+\mathrm{cos}\theta \simeq 2$ 
but neglecting the interband contribution 
%$1-\mathrm{cos}\theta \simeq 0$ 
as this is smaller than the interband term by an order
 $\mathcal{O}(q^2/k_{F0}^2)$. Eq.~(\ref{eq12}) is then evaluated as
\begin{eqnarray}
&&\sum_{\bm{p}} \mathrm{cos}^2\phi_{\bm{p}}\left(\frac{p}{m}\right)^2
\nonumber \\
&&\mathrm{tr}\left\{\mathrm{Im}\left\{U_{+}\left[(\tilde{G}^{R})^2-(\tilde{G}^{A})^2\right]U_{+}^{*}\right\}\tilde{G}^{A}\tilde{G}^{R}\right\}
\nonumber \\
&&=
\frac{2\pi}{m}\left(\frac{\tau}{\hbar}\right)^2\left(\nu_{+}+\nu_{-}\right),
\label{eq14} \\
\nonumber \\
&&\sum_{\bm{p}}\mathrm{cos}^2\phi_{\bm{p}}\alpha \frac{p}{m}
\nonumber \\
&&\mathrm{tr}\left\{\mathrm{Im}\left\{U_{+}\sigma_z\left[(\tilde{G}^{R})^2-(\tilde{G}^{A})^2\right]U_{+}^{*}\right\}\tilde{G}^{A}\tilde{G}^{R}\right\}
\nonumber \\
&&=
2\pi\alpha\frac{\tau^2}{\hbar^3}\left(\frac{\nu_{+}}{k_{F+}}-\frac{\nu_{-}}{k_{F-}}\right).
\label{eq15}
\end{eqnarray}

We note that it is crucial, as emphasized in Ref.~\cite{Oreg}, that in
evaluating the above integrals Eqs.~(\ref{eq14})-(\ref{eq15}) the
asymmetry between electron and hole spectra has to be taken into account, for
otherwise electron and hole drag would compensate each other
completely, yielding a null result. Using Eqs.~(\ref{eq42})-(\ref{eq43}) for $k_{\pm}$ and $\nu_{\pm}$, we
find that the SO coupling term Eq.~(\ref{eq15}) vanishes and the
nonlinear susceptibility Eq.~(\ref{eq12}), in the diffusive limit, remains unchanged in the
presence of Rashba SO coupling:
\begin{equation}
\Gamma_x(q,\omega) = -\frac{2\omega q}{\pi}
\left(\frac{\tau}{\hbar}\right)^2.
\label{eq16}
\end{equation}
In the above calculation we have only taken into account the diffuson
vertex correction to the charge current vertex. In the following
we include also the diffuson pole correction to the
charge density vertex, whereby Eq.~(\ref{eq12}) becomes
\begin{eqnarray}
&&\Gamma_x(\bm{q},\omega) \nonumber \\
&=& -\frac{\omega q}{\pi}\sum_p \mathrm{cos}^2 \phi_p\left(\frac{p}{m}\right)
\mathrm{tr}\left\{
  \mathrm{Im}\left\{U_{+}\left(\frac{p}{m}-\alpha\sigma_z\right) \right.\right.\nonumber \\
&&\left.\left.\left[\frac{1}{Dq^2+i\omega}(\tilde{G}^{R})^2-\frac{1}{Dq^2-i\omega}(\tilde{G}^{A})^2\right]U_{+}^{*}\right\}\right\}. \nonumber \\
%&&\left.\left.-\alpha\mathrm{Im}\left\{U_{+}\sigma_z\left[\frac{1}{Dq^2+i\omega}(\tilde{G}^{R})^2-\frac{1}{Dq^2-i\omega}(\tilde{G}^{A})^2\right]U_{+}^{*}\right\}\right\}\tilde{G}^{A}\tilde{G}^{R}\right\}.
\label{eq17}
\end{eqnarray}
Evaluating the integrals, we find the complete expression of the nonlinear susceptiblity
in the diffusive limit with Rashba SO coupling
\begin{equation}
\Gamma_x(q,\omega) = -4\frac{\omega Dq^2}{\left(Dq^2\right)^2+\omega^2}
\frac{\nu Dq}{\varepsilon_{F0}},
\label{eq18}
\end{equation}
which is unmodified in the presence of SO coupling. This rather peculiar
result for the Rashba SO coupling can be explained in light of the
following. First, the diffuson vertex renormalization restores
the charge current to its bare value $\bm{p}/m$ without SO coupling. Second,
it will be recalled that, without
SO coupling, the nonlinear susceptibility
Eq.~(\ref{eq8}) in the diffusive limit is commensurate with the
imaginary part of the irreducible polarizability only to the
lowest order of $ql$ and $\omega\tau$ \cite{Oreg,Jauho}. We find 
that the same statement holds true in the presence of SO coupling. 
The polarizability in the presence of disorder is determined as
\begin{eqnarray}
\Pi &=& \mathrm{tr}\int\frac{\mathrm{d}\varepsilon}{2\pi
  i}\sum_{\bm{p}} \nonumber \\
&&\left\{n_F(\varepsilon)\left[G^R_{\varepsilon+\omega}\left(\bm{k}+\bm{q}\right)G^R_{\varepsilon}\left(\bm{k}\right)-G^A_{\varepsilon+\omega}\left(\bm{k}+\bm{q}\right)G^A_{\varepsilon}\left(\bm{k}\right)\right]
  \right. \nonumber \\
&&\left. \left[n_F(\varepsilon+\omega)-n_F(\varepsilon)\right]\frac{1}{\tau(Dq^2-i\omega)}
G^R_{\varepsilon+\omega}\left(\bm{k}+\bm{q}\right)G^A_{\varepsilon}\left(\bm{k}\right)\right\}
  \nonumber \\
&\simeq& 2\nu\left\{1+\frac{i\omega}{Dq^2-i\omega}\left[1-\frac{(ql)^2}{2}(1-\gamma^2)\right]\right\},
\label{eq41}
\end{eqnarray}
where it is seen that the leading term is unmodified in the
presence of SO interaction, and the SO coupling correction comes in only in
the second order $(ql)^2$. It follows that the polarizability
$\Pi(q,\omega)$ and therefore the nonlinear susceptiblilty
$\Gamma(q,\omega)$ remain unchanged to the lowest order in $ql$ and
$\omega\tau$ in the presence of SO coupling. It follows that, in the drag
resistivity Eq.~(\ref{eq4}), since
the dominant contribution to the momentum integral comes from small $q$
at low temperatures, $\rho_{\mathrm{D}}$ also remains
 unchanged in the presence of SO coupling, i.e. the SO
coupling correction to the Coulomb drag becomes completely suppressed in the presence of disorder.

\subsection{Drag Resistivity}

By solving the Dyson equation for the the interlayer
potential in the random phase approximation, $U_{12}$ in
Eq.~(\ref{eq4}) is given as \cite{Oreg}
\begin{equation}
U_{12}\left(q,\omega\right) = \frac{q}{4\pi e^2 \mathrm{sinh}(qd) \Pi_1\left(q,\omega\right) \Pi_2\left(q,\omega\right)},
\label{eq26}
\end{equation}
where $d$ is the interlayer spacing; $\Pi\left(q,\omega\right)$ is the
polarizability Eq.~(\ref{eq23}) for the
ballistic case or Eq.~(\ref{eq41}) for the diffusive case. In the
low-energy long-wavelength limit $u, x \ll 1$, the expressions $\Pi \simeq 2\nu$ or $\Pi \simeq
2\nu Dq^2/(Dq^2-i\omega)$ for the ballistic or the diffusive regime
remain unchanged in the presence of SO coupling, from which it follows that the expression of the interlayer
potential Eq.~(\ref{eq26}) also remains the same within the
assumptions made.

The drag resistivity $\rho_{\mathrm{d}}$ in the diffusive regime and
$\rho_{\mathrm{b}}$ in the ballistic regime for two identical 2DEG layers without
spin-orbit coupling are well known \cite{Oreg,Jauho2,MacDonald}:
\begin{eqnarray}
\rho_{\mathrm{d}} &=&
\frac{\hbar}{e^2}\frac{\pi^2}{12}\left(\frac{k_B
    T}{\epsilon_{F0}}\right)^2\mathrm{ln}\frac{T_0}{2T}
\frac{1}{\left(q_{\mathrm{TF}}d\right)^2\left(k_{F0} l\right)^2},
\label{eq28} \\
\rho_{\mathrm{b}} &=&
\frac{\hbar}{e^2}\frac{\pi^2\xi\left(3\right)}{16}\left(\frac{k_B T}{\epsilon_{F0}}\right)^2\frac{1}{\left(q_{\mathrm{TF}}d\right)^2\left(k_{F0} d\right)^2},
\label{eq29}
\end{eqnarray}
where $q_{\mathrm{TF}} = 4\pi e^2\nu$ is the Thomas-Fermi wavenumber and $k_B T_0 = \hbar D q_{\mathrm{TF}}/d$. In the presence of electron-electron interaction, the spin-orbit coupling
strength $\alpha$ will be renormalized, becoming a function of the interaction parameter $r_s = \sqrt{2}e^2
m/\hbar^2 k_{F0}$. The detailed evaluation of the many-body
renormalization of the SO coupling is beyond the scope of this paper,
nevertheless we note that in the statically screened case the
renormalization of the SO coupling $\alpha$ amounts simply to the mass
renormalization \cite{Raikh} $\alpha(m/m^{*})$, which however may not be true
if dynamical screening is taken into account. Combining Eqs.~(\ref{eq4}), (\ref{eq25}) and (\ref{eq18}), we immediately
arrive at the results that: (1) the drag resistivity is unchanged in the
diffusive limit $\rho_{\mathrm{D}} = \rho_{\mathrm{d}}$; (2) the drag resistivity in the ballistic limit is
enhanced as
\begin{eqnarray}
\rho_{\mathrm{D}} &=& \rho_{\mathrm{b}} \left[1+\left(\gamma^*\right)^2\right],
\label{eq31}
\end{eqnarray}
where we have denoted the renormalized SO coupling strength as
$\gamma^* = \gamma(m/m^{*})$ within the static screening approximation
made.

\section{Spin Coulomb Drag}

In the following we perform a calculation of the spin Coulomb drag
resistivity in the ballistic limit. Analogous to Eq.~(\ref{eq4}) the spin Coulomb drag resistivity is
given as
\begin{eqnarray}
\rho_{\uparrow\downarrow} &=& \frac{\hbar^2}{16\pi k_B T\sigma_{\uparrow}\sigma_{\downarrow}}\sum_{\bm{q}}\int_{0}^{\infty}
\frac{\mathrm{d}\omega}{\mathrm{sinh}^2\left(\hbar\omega/2k_B
    T\right)} \nonumber \\
&&\Gamma_{\uparrow x}\left(q,\omega\right)\Gamma_{\downarrow x}\left(q,\omega\right)\left\vert U\left(q,\omega\right)\right\vert^2,
\label{eq32}
\end{eqnarray}
where $U$ is the Coulomb potential, $\sigma_{\uparrow,\downarrow}$ is
respectively the Boltzmann conductivity for spin up and spin down carriers.

First we consider the case without SO coupling. The spin-up
and spin-down current are given by
\begin{eqnarray}
J_{\uparrow x} &=& \frac{\hbar k_x}{2m}
\left[
\begin{array}{cc}
1 & 0 \\
0 & 0
\end{array}
\right], \label{eq33} \\
J_{\downarrow x} &=& \frac{\hbar k_x}{2m}
\left[
\begin{array}{cc}
0 & 0 \\
0 & 1
\end{array}
\right]. \label{eq34}
\end{eqnarray}
Using these expressions in Eq.~(\ref{eq19}) yields
\begin{eqnarray}
{\Gamma}_{\uparrow x}(\bm{q},\omega)
&=&\frac{q \tau}{m}\sum_{\bm{p}}
\mathrm{Im}\left\{\frac{n_F(\epsilon_{\bm{p}\uparrow})-n_F(\epsilon_{\bm{p+q}\uparrow})}{\omega+\epsilon_{\bm{p}\uparrow}-\epsilon_{\bm{p+q}\uparrow}+i0^{+}}\right\},
\nonumber \\
&=& -\frac{q \tau}{m}\mathrm{Im}\Pi_{\uparrow}(\bm{q},\omega)
\label{eq35}
\end{eqnarray}
and similarly for down spin. Putting into Eq.~(\ref{eq32}) gives the
expression for the spin drag resistivity without SO coupling
\begin{eqnarray}
\rho_{\uparrow\downarrow} &=& \frac{\hbar^2}{16\pi e^2 k_B T n_{\uparrow}n_{\downarrow}}\sum_{\bm{q}}\int_{0}^{\infty}
\frac{\mathrm{d}\omega}{\mathrm{sinh}^2\left(\hbar\omega/2k_B
    T\right)} \nonumber \\
&&\mathrm{Im}\Pi_{\uparrow}\mathrm{Im}\Pi_{\downarrow}\left\vert U\right\vert^2,
\label{eq36}
\end{eqnarray}
consistent with Refs.~\cite{Vignale1,Flens,Vignale2}. We note however
that our result is smaller by a factor of four because in our definition of the spin up and spin
down currents Eqs.~(\ref{eq33})-(\ref{eq34}) we have included a factor
of $1/2$ for spin-$1/2$.

We now consider spin Coulomb drag in the presence of Rashba SO
coupling. Since in the ballistic limit the spin current with Rashba SO
coupling is diagonal in spin space $J_{sx} = (\hbar
k_x/2m)\sigma_z$, the spin up and spin down currents are still given by Eqs.~(\ref{eq33})-(\ref{eq34}).
Transforming them into the chiral basis and substituting the resulting
expressions into
Eq.~(\ref{eq19}) for the nonlinear susceptibility, it can be shown that, similar to Eq.~(\ref{eq21}),
in the ballistic limit the off-diagonal terms of the currents
$\tilde{J}_{\uparrow x}$, $\tilde{J}_{\downarrow x}$ do not contribute and the nonlinear
susceptibility for $\Gamma_{\uparrow x}$ or $\Gamma_{\downarrow x}$
still maintains the same form given by Eq.~(\ref{eq22}), where now $\tilde{J}_{11,22} = \hbar
k_x/4m$. Evaluating the momentum integral then gives
\begin{eqnarray}
\Gamma_{\uparrow,\downarrow x}\left(\bm{q},\omega\right) &=& -\frac{\nu D
  q}{\epsilon_{F0}}\frac{u}{\left(1-\gamma^2\right)^{3/2}} \nonumber \\
&\simeq& -\frac{\nu D
  q}{\epsilon_{F0}}u\left(1+\frac{3}{2}\gamma^2\right).
\label{eq37}
\end{eqnarray}
\\
For the Coulomb potential, we can use in the low temperature regime
the statically screened expression, since by virtue of Eq.~(\ref{eq27}) the
real part of the polarizability is unchanged by SO coupling, the Coulomb potential is
simply given by
\begin{equation}
U\left(q\right) = \frac{2\pi e^2}{q+q_{\mathrm{TF}}}.
\label{eq38}
\end{equation}
Putting Eqs.~(\ref{eq37})-(\ref{eq38}) into the Eq.~(\ref{eq32}), we find in a paramagnetic
system $n_{\uparrow} = n_{\downarrow} = n/2$ the spin drag resistivity
\begin{eqnarray}
\rho_{\uparrow\downarrow} &=&
\frac{\hbar}{e^2}\frac{\pi}{3}\left[1+3\left(\gamma^*\right)^2\right]\left(\frac{q_{\mathrm{TF}}}{2k_{F0}}\right)^2
\left(\frac{k_B T}{\epsilon_{F0}}\right)^2 \label{eq39} \\
&&\int_0^{\infty}\mathrm{d}q
\frac{q}{\left(q+q_{\mathrm{TF}}\right)^2} \nonumber \\
&\simeq& \frac{\hbar}{e^2}\frac{\pi}{3}\left[1+3\left(\gamma^*\right)^2\right]\left(\frac{q_{\mathrm{TF}}}{2k_{F0}}\right)^2
\left(\frac{k_B T}{\epsilon_{F0}}\right)^2 \nonumber \\
&&\;\;\;\left[\mathrm{ln}\left(1+\frac{2k_{F0}}{q_{\mathrm{TF}}}\right)-\frac{1}{1+q_{\mathrm{TF}}/2k_{F0}}\right], \nonumber
\end{eqnarray}
where in the last line in order to evaluate the divergent $q$
integral we have introduced an upper cutoff at $q = 2k_{F0}$. Our
result Eq.~(\ref{eq39}) is in agreement with Ref.~\cite{Flens} 
in the absence of SO coupling. We observe here that the presence of the SO interaction also enhances the spin
Coulomb drag.

We comment in passing that, in the diffusive regime, correlated impurity
scattering has to be taken into account since the spin-up and
spin-down electrons are scattered by the same set of impurities in the
sample, as emphasized in
Ref.~\cite{Vignale1}. In this paper we have not attempted
to include the diffusive limit as the effect of
correlated impurity scattering requires taking into account a
different set of diagrams \cite{Gornyi} and is beyond the scope of the current paper.
However, taking account of such correlated impurity scattering effect
is expected to further enhance the spin Coulomb drag in the presence
of SO coupling also. \\

\section{Conclusion}

Before concluding, we point out that our theory indicates rather small
effect of SO coupling on the Coulomb and spin drag, at least in the
weak SO coupling regime considered in our work. The drag is either
unaffected by the SO coupling or is affected only in the quadratic
order of the SO coupling strength. So the signature of SO coupling may
not be easy to discern in experimental drag measurements. On the other
hand, we do find that the SO coupling enhances the bilayer Coulomb
drag by the factor $\left[1+\left(\gamma^*\right)^2\right]$ in the
ballistic regime, indicating that Coulomb drag would be enhanced by
the presence of the SO coupling. This is actually consistent with the
available experimental findings. In particular, the measured bilayer
hole drag \cite{HoleD1,HoleD3} seems to be larger than the
corresponding theoretical results \cite{HoleD2,HoleD4} even after the
inclusion of many-body effects. Since the SO coupling is strong in
GaAs holes, the dimensionless SO coupling strength
$\left(\gamma^*\right)^2$ could be large, particularly at the low
experimental hole densities (because the definition of $\gamma$
involves the Fermi velocity in the denominator). This could lead to an
appreciable enhancement of the 2D Coulomb drag for low density holes
in the ballistic regime bringing experiment and theory closer
together. Unfortunately, we cannot make any quantitative comments on
this issue since our theory is explicitly a weak SO coupling
expansion. 

In summary, we have calculated the Coulomb drag resistivity of a
double-layer 2DEG system including spin-orbit coupling. In the
diffusive limit we find that there is no change in the Coulomb drag
resistivity while in the ballistic limit we find the 
drag resistivity to be enhanced by a factor of $\left[1+\left(\gamma^*\right)^2\right]$.
We also apply the formalism to the problem of spin Coulomb drag
and find that the spin drag resistivity in the ballistic limit is
similarly enhanced as well. These
enhancement effects in the ballistic regime are due to the SO-induced enhancement to the
nonlinear susceptibility in both the charge Coulomb drag and spin
Coulomb drag.

\section{acknowledgement}

We gratefully acknowledge useful discussions with Euyheon Hwang. This
work is supported by NSF, LPS, and ONR.


\begin{thebibliography}{18}
\bibitem{remark} This holds only under the assumption that the mean free time $\tau$ is a
  constant independent of momentum, as is the case considered throughout
  our paper.
\end{thebibliography}
\end{document}